\documentclass[journal=jacsat,manuscript=article]{achemso}

\usepackage[version=3]{mhchem} 

\usepackage{xcolor} 
\usepackage[hidelinks]{hyperref}
\usepackage{pdfpages}

\usepackage{etoolbox}

\makeatletter
\renewcommand*{\acs@author@fnsymbol@symbol}[1]{
    \ifcase #1 *\or
    1\or
    2\or
    3\or
    4\or
    5\or
    6\or
    7\or
    8\or
    9\or
    10
    \fi
}
        
\renewcommand*\acs@contact@details{
    {\sffamily *\,Email: \acs@email@list }%
    \acs@number@list
}           

\patchcmd{\acs@address@list@auxii}
{\acs@author@fnsymbol{\acs@affil@marker@cnt}}
{\textsuperscript{\acs@author@fnsymbol{\acs@affil@marker@cnt}}}
{}{}

\patchcmd{\acs@address@list@auxii}
{{\acs@author@fnsymbol{\acs@affil@marker@cnt}\@nameuse{@altaffil@\@roman\@tempcnta}\par}}
{{\textsuperscript{\acs@author@fnsymbol{\acs@affil@marker@cnt}}\@nameuse{@altaffil@\@roman\@tempcnta}\par}}
{}{}        

\makeatother    

\author{Seyed Khalil Alavi}
\affiliation[fmq]
{Institute for Functional Matter and Quantum Technologies, University of Stuttgart, 70569 Stuttgart,
Germany}
\alsoaffiliation[iqst]
{Center for Integrated Quantum Science and Technology (IQST), University of Stuttgart, 70569 Stuttgart,
Germany}

\author{Zijie Sheng}
\affiliation[fmq]
{Institute for Functional Matter and Quantum Technologies, University of Stuttgart, 70569 Stuttgart,
Germany}
\alsoaffiliation[iqst]
{Center for Integrated Quantum Science and Technology (IQST), University of Stuttgart, 70569 Stuttgart,
Germany}

\author{Haneul Lee}
\affiliation[kaist]
{Department of Physics, Korea Advanced Institute of Science and Technology (KAIST), Daejeon 34141,Republic of Korea}

\author{Hansuek Lee}
\affiliation[kaist]
{Department of Physics, Korea Advanced Institute of Science and Technology (KAIST), Daejeon 34141,Republic of Korea}

\alsoaffiliation[GS]{Graduate School of Quantum Science and Technology, KAIST,
Daejeon 34141, Republic of Korea
}

\author{Sungkun Hong}
\affiliation[fmq]
{Institute for Functional Matter and Quantum Technologies, University of Stuttgart, 70569 Stuttgart,
Germany}

\alsoaffiliation[iqst]
{Center for Integrated Quantum Science and Technology (IQST), University of Stuttgart, 70569 Stuttgart,
Germany}
\email{sungkun.hong@fmq.uni-stuttgart.de}
\title[An \textsf{achemso} demo]
  {Enhanced optomechanical coupling between an optically levitated particle and an ultrahigh Q optical microcavity}

\setkeys{acs}{keywords = true}

\abbreviations{}

\begin{document}

\begin{abstract}
Exploring the dynamics of an optically levitated dielectric micro- and nanoparticle is an exciting new subject in quantum science. Recent years have witnessed rapid advancements in attaining quantum-limited optical detection and control of a nanoscale particle by coupling its motion to a high-finesse optical cavity in the resolved-sideband regime. In order to control the particle deeper in the quantum regime, it is necessary to significantly enhance the coupling between the particle and the cavity. Here, we present a novel platform that can allow for achieving this. Our system consists of a conventional optical tweezer and a toroidal optical microcavity with an ultrahigh quality (Q) factor. The optomechanical coupling between the particle and the cavity is established by placing the particle in the near field of the cavity. The significantly reduced mode volume allows us to achieve a 50-fold increase in the single photon optomechanical coupling compared to a conventional Fabry-Pérot cavity with macroscopic mirrors, while ultralow loss of the cavity brings the system close to the resolved-sideband regime. Our approach paves the way for enabling quantum experiments on levitated mesoscopic particles with high quantum cooperativity near the resolved-sideband regime.

\end{abstract}

\section{Introduction}
Studying the dynamics of optically levitated particles has emerged as a flourishing field in quantum optomechanics, offering a unique platform for exploring fundamental aspects of quantum physics and advancing quantum technologies \cite{gonzalez-ballestero_levitodynamics_2021}. Investigating the quantum behavior of levitated particles, for instance, can provide insights into fundamental principles of quantum physics such as quantum superposition and entanglement in larger mass and length scales \cite{romero-isart_toward_2010, chang_cavity_2010, romero-isart_large_2011, scala_matter-wave_2013, yin_large_2013}. Recent research has also shown great promise of levitated particles as precision sensors with a wide range of applications from force and acceleration sensing \cite{ranjit_zeptonewton_2016, ahn_ultrasensitive_2020, monteiro_force_2020} to gravitational wave detection \cite{aggarwal_searching_2022}.%

In the last decade, the field has made remarkable progress in controlling the motion of levitated particles in the quantum regime. A notable experimental method that has been developed during this period is the use of high-Q optical cavities \cite{meyer_resolved-sideband_2019, delic_levitated_2020}, which provide significantly enhanced light-matter interactions and, therefore, optical readout and control. In particular, optical cavities in the resolved-sideband regime - where the cavity's loss rate ($\kappa$) is comparable to or smaller than the motional frequency of the particle ($\Omega$) - allow the selective enhancement of inelastic scattering processes (Stokes and anti-Stokes) between the particle and the light through laser frequency detuning, thus enabling various quantum optical operations such as beam splitter and two-mode squeezing interactions \cite{aspelmeyer_cavity_2014}. Recent breakthroughs achieved with the cavities in the resolved-sideband regime include the preparation of the particle near the quantum ground state in one \cite{delic_cooling_2020} and two translational directions \cite{piotrowski_simultaneous_2023}, the simultaneous cooling of rotational motions \cite{pontin_simultaneous_2023}, and the realization of long-range interactions between two levitated particles \cite{vijayan_cavity-mediated_2024}.

To perform experiments deeper in the quantum regime, it is necessary to further increase the optomechanical coupling between the particle and the cavity. However, this remains a challenging task due to limitations in the parameters offered by conventional Fabry-Pérot cavities with typical sizes on the order of millimeters. A promising approach to overcoming this challenge is to employ cavities with micron-scale mode volumes, which can dramatically increase the optomechanical coupling by several orders of magnitude compared to cavities with macroscopic volumes, provided that they can be driven with a similar level of laser input power. This approach has recently been realized with microfabricated photonic crystal cavities \cite{magrini_near-field_2018}.  However, the achieved cavity loss rate was four orders of magnitude larger than the mechanical frequency of the particle, thereby strictly preventing access to the resolved-sideband regime, and the demonstrated parameters have also not yet reached the level required for deep quantum regime experiments.

In this study, we utilize a toroidal optical microcavity fabricated from thermally grown silica to develop a novel cavity optomechanical system with a levitated nanoparticle. The silica microtoroid cavity exhibits ultralow optical loss and high power handling due to its exceptional material and structural qualities. Together with the micron-scale mode volume, it constitutes a promising microcavity for realizing a hybrid optomechanical system near the resolved-sideband regime with exceptional quantum cooperativity. By placing the particle approximately $300~nm$ from the surface of the microtoroid cavity, we experimentally observe strong and stable optomechanical coupling signals. The achieved single-photon coupling strength is $g_0⁄2\pi  = 14.7 \pm 0.1~Hz$, surpassing the previously reported value obtained with macroscopic cavities by two orders of magnitude \cite{meyer_resolved-sideband_2019, delic_levitated_2020}. Moreover, the observed cavity loss rate of $\kappa⁄2\pi = 4.4~MHz$ is three orders of magnitude smaller than that of the microcavity used in the previous study \cite{magrini_near-field_2018}, bringing our system a factor of ten from the resolved-sideband regime. With a realistic improvement of the devices and the vacuum environment, we envisage that our system can realize quantum cooperativity far beyond unity near the resolved-sideband regime, opening up a new avenue for quantum optomechanics experiments with levitated nanoparticles.



\begin{figure}[t!]
    \centering
    
\includegraphics{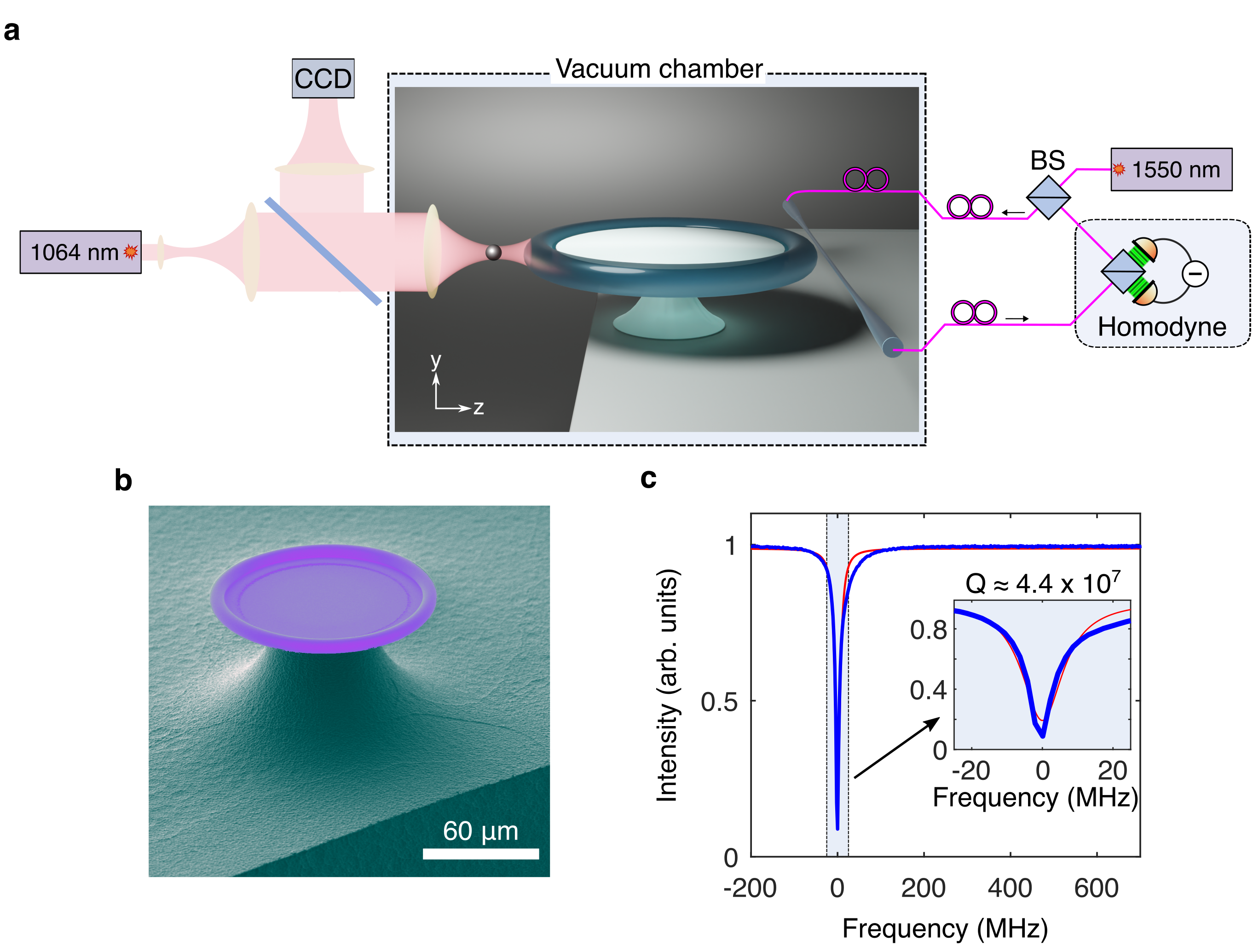}
    \caption{ \textbf{Mictocavity-levitated optomechanics}. \textbf{a}, Schematic of the experimental setup. A $1064~nm$ laser beam is focused by a $0.95$ NA microscope objective inside a vacuum chamber, creating an optical tweezer that traps a single silica nanoparticle with a diameter of $150~nm$. Following the trapping of a particle, a toroidal silica microcavity is moved towards the particle. It is oriented so that the equator of the cavity faces the focal point of the objective. A tapered fiber is employed to pump the microcavity with a $1550~nm$ laser. The pumped cavity field is coupled back to the fiber and transmitted to a homodyne detection setup to measure optomechanical coupling between the cavity field and the center of mass motion of the particle. \textbf{b}, Scanning electron microscope image of a silica microcavity fabricated near the edge of a silicon substrate. \textbf{c}, Response of the microcavity to the pump laser. The laser wavelength is scanned around the resonance of the fundamental mode of the cavity ($1550.06~nm$) with an optical loss rate of $\kappa⁄2\pi = 4.4~MHz$, corresponding to a quality factor of $Q \approx 4.4 \times 10^7$. } 
    \label{fig1}
\end{figure}



\section{Results}
Our experimental setup consists of an optical tweezer for trapping a single silica nanoparticle and a toroidal silica microcavity, both located in a vacuum chamber (see Fig. \ref{fig1}a). The optical tweezer is formed using a microscope objective with a numerical aperture (NA) of $0.95$ that focuses a laser beam with a wavelength of $1064~nm$ and a power of $\approx 450~mW$. The objective is aligned so that its focal point faces the equator of the toroidal microcavity. The silica microcavities are fabricated on a silicon chip through a series of lithography and etching processes, followed by $CO_2$ laser reflow \cite{armani_ultra-high-q_2003} (see also Supporting Information). Due to the extremely low material loss of the silica film and the molecular-scale surface roughness achieved by the reflow process, these microcavities typically exhibit Q factors in excess of 100 million \cite{anetsberger_near-field_2009}. The toroidal microcavity has a major diameter of $70~\mu m$, and a minor diameter of $7~\mu m$, and is within $100~\mu m$ from the edge of a silicon chip, which allows us to access the circumference of the toroid with the tweezer beam from the side (see Fig. \ref{fig1}b). The fundamental mode of the cavity has a resonance at a wavelength of $1550.06~nm$ and an optical Q factor of $4.4 \times 10^7$ (see Fig. \ref{fig1}c). A tapered optical fiber with a transmission efficiency of $\eta_t = 0.8$  is used to couple the laser light into and out of the microcavity \cite{spillane_ideality_2003, cai_observation_2000}. Together with a fiber-to-cavity coupling efficiency of $\eta_{cav} = 0.81$, and accounting for other losses in the setup, it yields a total detection efficiency of $0.15$. The tapered fiber and the microcavity can be moved independently with respect to the optical tweezer using three-axis nanopositioners. The microcavity position is primarily monitored during the alignment using the tweezer microscope objective in imaging mode (Fig. \ref{fig1}a). Not shown in the figure is a separate imaging module installed perpendicular to the tweezer axis, which provides an additional view of the microcavity and the fiber and aids the alignment process.

\begin{figure}[t!]
    \centering
    \includegraphics{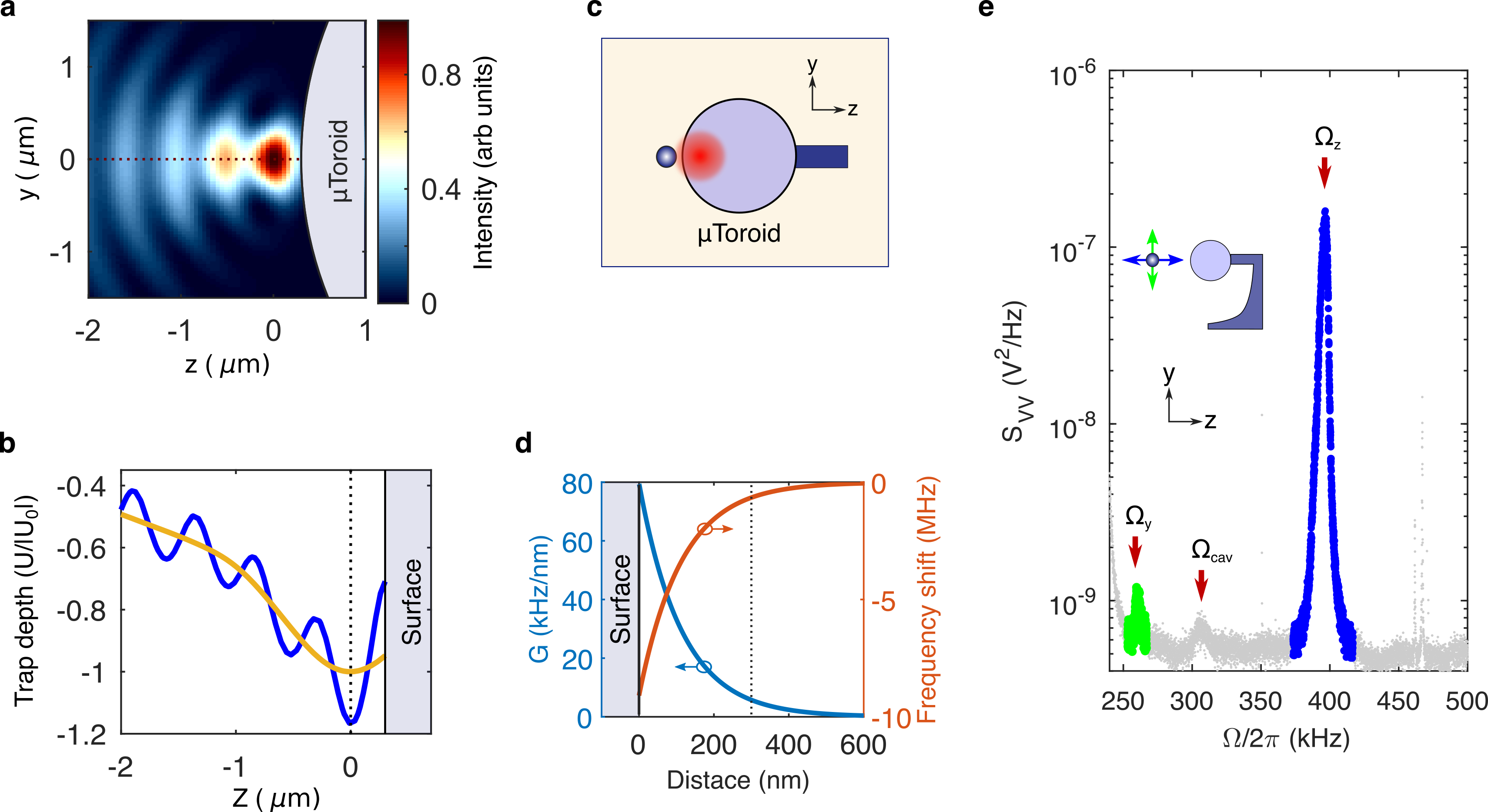}
    \caption{\textbf{Optomechanical coupling}. \textbf{a}, Simulation of the tweezer beam incident on the equator of the microtoroid. The panel shows the light field intensity in the plane parallel to the $yz$-plane defined in Fig. \ref{fig1}a, intersecting the middle of the tweezer beam along the beam propagation axis. The cross-sectional profile of the microtoroid ($\mu$Toroid) is also shown. The forward propagating tweezer field interferes with its reflection from the toroid surface, producing the modulated intensity pattern. The intensity maximum is formed 300 nm away from the surface. \textbf{b}, Resulting optical potential profile (blue) along the tweezer beam axis (the dotted line in a) near the surface of the microtoroid. The profile of the unmodified potential in the absence of the toroid (yellow) is also shown for comparison. Both profiles are normalized with the potential depth of the unmodified potential. The interference results in multiple potential wells with widths much narrower than the unmodified potential, thus enhancing the frequency of the particle motion along the $z$-direction. \textbf{c}, Schematic of the near-field coupling between the nanoparticle and the cavity. A nanoparticle is located at a close distance to the microcavity, where the evanescent portion of the fundamental mode (shaded with red) of the cavity exists. \textbf{d}, Numerical calculation of the cavity's resonance frequency shift and its derivative ($G$) resulted from the presence of a particle at varying distances from the surface of the microtoroid. The dotted line indicates the position of the particle in our platform ($300~nm$), with an estimated $G^{sim}$ of $5.8~kHz/nm$. \textbf{e}, Measured power spectral density of the homodyne signal from the cavity. The thermal motion of the particle induces cavity frequency fluctuation, showing a peak corresponding to the center of mass motion of the particle along $z$ at a frequency of $\Omega_z⁄2\pi = 396.6~kHz$ (blue). Here, the particle’s position is adjusted by steering the tweezer beam to maximize the peak at the frequency of $\Omega_z⁄2\pi$, thus the coupling. The particle's motion along y is faintly coupled to the cavity ($\Omega_y⁄2\pi = 259.6~kHz$, green). A peak at around the frequency $\Omega_{cav}/2\pi  \approx 306~kHz$ results from the mechanical vibration of the tapered fiber relative to the cavity.}
    \label{fig2}
\end{figure}

To perform the experiment, we first trap a silica nanoparticle (diameter $150~nm$) at a pressure of around $200~mbar$. The pressure is subsequently reduced to a fraction of a $mbar$. All the measurements presented in this study were performed at $0.36 ~mbar$. The microcavity is then moved towards the focal plane of the tweezer objective. When the tweezer beam is exposed to the microtoroid, it results in a thermal drift of the sample upon absorption of light. However, this drift is stabilized when the sample reaches the thermal equilibrium. It typically takes about fifteen minutes once the microcavity is brought within tens of micrometers from the focal plane. Approaching the focal plane leads to an adiabatic transformation of the tweezer-induced optical trap from a single Gaussian potential to multiple potential wells (see Fig. \ref{fig2}a and b). This is due to the formation of standing waves resulting from the reflection of the tweezer beam at the surface of the microcavity \cite{magrini_near-field_2018, ju_near-field_2023-1, diehl_optical_2018}. The locations of these potential wells are solely determined by the properties of the reflecting surface and the wavelength of the tweezer field. In our case, the nearest well is $300~nm$ away from the surface of the microcavity (see Fig. \ref{fig2}a and b and Supporting Information). We load the particle into the nearest well by slowly approaching the microcavity toward the focal plane of the unperturbed tweezer trap, thus successively pushing the particle from one potential well to its neighbor closer to the microtoroid\cite{magrini_near-field_2018, diehl_optical_2018}. We also note that the potential wells impose a strong confinement along the tweezer axis, $z$, which leads to an increase in the frequency of the particle’s oscillation in the trap (see also Supporting Information, Fig. S4). In our experiment, we observe a significant increase in the particle's mechanical frequency along $z$ from $\Omega_z⁄2\pi = 80.9~kHz$ to $\Omega_z⁄2\pi = 396.6~kHz$ at the nearest well of the standing wave trap. This leads to the cavity loss to the particle frequency ratio of $\kappa/\Omega_z=11$, placing our system closer to the resolved-sideband regime.

 
\begin{figure}[tp]
    \centering
\includegraphics{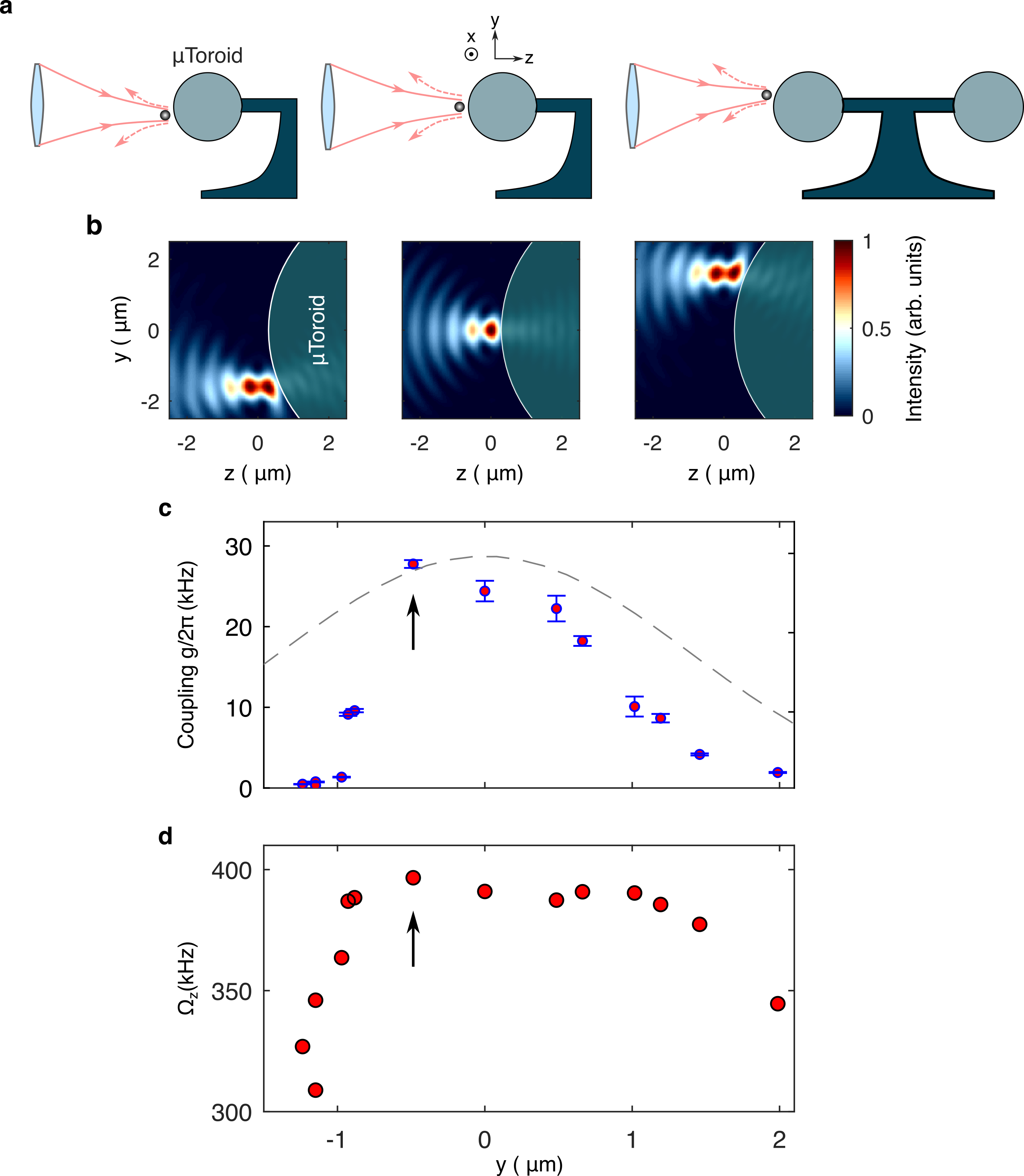}
    \caption{\textbf{Tuning of optomechanical coupling and particle’s oscillation frequency}. \textbf{a}, Beam steering illustration. The tweezer beam is steered along the y-axis with the particle positioned in the closest potential well. Simultaneously, we record the cavity response. \textbf{b}, Simulation of the trapping beam when the focus is away from the equator. When the beam focus is steered away from the cavity’s midpoint ($y = 0$), the standing waves deform and exhibit asymmetry. The left (right) panel displays the tweezer beam when the focus is at $y = -1.6~\mu m$ ($y= 1.6~\mu m$). \textbf{c}, Optomechanical coupling for the experiment described in (a). The dashed line is proportional to the intensity of the cavity field, $ \propto \vert E(y,z)\vert^2 $, at $z = 300~nm$, which is numerically simulated assuming the cross-sectional profile of the microtoroid is a perfect circle. It serves as a guide for the eye. The PSD shown in Fig.~\ref{fig2}e is obtained at the y-position indicated by the arrows here and in panel d. The coupling is maximized near the equator, where the cavity field is most intense. The non-ideal symmetry of the cavity’s geometry and the onset of thermal drift triggered by the beam steering contribute to an asymmetry observed in both the coupling and the frequency shifts shown in panel d. \textbf{d}, Mechanical frequency of the particle’s motion along $z$ for the same experiment as in (b). The significant change in the particle’s frequency is also observed since the beam steering deforms the standing wave formed by the tweezer beam and its reflection from the cavity surface.}
    \label{fig:3}
\end{figure}



The presence of a dielectric nanoparticle at a short distance from the surface of a microcavity leads to an increase in the effective refractive index sampled by the exponentially decaying evanescent field (see Fig. \ref{fig2}c and d). Consequently, the displacement of the particle, $\delta \zeta$ leads to a shift in the resonance frequency of the cavity, $\omega_c (\zeta + \delta \zeta)- \omega_c ( \zeta)  \approx -G\delta \zeta$, where $\zeta = x,y,z$, indicates the direction of the motion. Here $G =  \partial \omega_c⁄\delta \zeta$ denotes the cavity frequency shift per particle displacement. The single-photon optomechanical coupling $g_0$ is then obtained by multiplying $G$ with the particle's zero-point fluctuation $x_{zpf} = (\hbar/ 2m \Omega_z)^{1/2}$, i.e., $g_0 =  x_{zpf} G$, where $m$ is the particle mass and $\Omega_z$ is the particle's oscillation frequency.

To determine the optomechanical coupling, we first position the nanoparticle near the microcavity's equator, where the fundamental optical mode of the cavity resides predominantly. Then, the small thermal motion of the particle causes the cavity's resonance frequency fluctuation, modulating the phase quadrature of the cavity's output field. We measure this by performing the homodyne measurement on the cavity output field while the pump laser is locked on resonance to the cavity’s fundamental mode. Figure \ref{fig2}e shows a power spectral density (PSD) of the measured homodyne signal while the motion of the particle is coupled to the cavity. 
The measured PSD is then converted to the PSD of the cavity frequency fluctuation, $S_{\omega \omega}(\Omega)$, using predetermined detection efficiency and laser input powers to the interferometer. We then deduce the coupling $G$ by comparing the power of the $S_{\omega \omega}(\Omega)$ around the particle's frequency and the particle's expected thermal fluctuation (see Supporting Information for more details).

From the measurement results, we obtain $G = 5.37 \pm 0.05~kHz/nm$, corresponding to a single-photon optomechanical coupling of $g_0⁄2\pi = 14.7 \pm 0.1~Hz$. This is in excellent agreement with the value expected from our finite element method simulation, $G^{sim} = 5.8~kHz/nm$, corresponding to $g^{sim}_0⁄2\pi = 15.9~Hz$ (see Fig. \ref{fig2}d and Supporting Information). This measurement is conducted with an input power of $12.5~\mu W$, from which we infer a light-enhanced optomechanical coupling of $g⁄2\pi = \sqrt{n_{cav}}  g_0⁄2\pi  = 27.8 \pm 0.4~kHz$ (with the number of cavity photons, $n_{cav} \approx~3.6 \times 10^6$).

Figure \ref{fig:3} illustrates the coupling of the particle’s motion to the fundamental mode of the microcavity as we change the particle's position along the $y$-axis. This is achieved by steering the tweezer beam while monitoring the cavity response. When the particle is displaced away from the equator plane (middle of the ring, as illustrated in  Fig. \ref{fig:3}a), both the coupling strength and the frequency of the particle’s motion along the $z$-axis decrease. The variations in coupling strength follow the spatial profile of the selected cavity mode, i.e., the fundamental mode, where its intensity is highest at the equator plane (see Fig. \ref{fig:3}c and Supporting Information). Consequently, we observe maximum optomechanical coupling at that location. The decrease of the particle’s frequency can be understood by observing the deformation of the optical potential wells and the reduction of the trap stiffness (see Fig. \ref{fig:3}b and d). 

Another intriguing aspect of our platform, in particular, compared to the previously studied photonic crystal cavities\cite{magrini_near-field_2018}, is the ability to harness the cavity's diverse mode spectrum. The microtoroid cavities employed in our experiment support multiple resonances with distinct spatial mode profiles, which also exhibit high quality factors comparable to the fundamental mode. In another set of experiments, we demonstrate optomechanical coupling with a higher mode of the cavity and observe a coupling strength comparable to that achieved for the fundamental mode (see Supporting Information, Fig. S7). This feature enables simultaneous optomechanical coupling to multiple cavity modes, providing a pathway to employ more sophisticated measurement and control schemes over the particle’s motion \cite{rossi_measurement-based_2018}.

\section{Conclusion}
In summary, we have introduced an ultrahigh Q toroidal microcavity as a promising optical interface for exploring the dynamics of optically levitated dielectric particles at the quantum limit. The small mode volume of the microcavity allowed us to obtain a single-photon optomechanical coupling two orders of magnitude larger than that of conventional Febry-P\'erot cavities \cite{delic_levitated_2020, meyer_resolved-sideband_2019}. By driving the cavity with the laser input power of $12.5~\mu W$, we achieved the enhanced optomechanical coupling to $g/2\pi = 27.8~kHz$. This would yield a quantum cooperativity $C_q = 4g^2⁄\kappa \gamma_{th} \sim 0.1$ at a pressure below $10^{-7}~mbar$, where the mechanical decoherence rate $\gamma_{th}$ is limited to the photon recoil by the tweezer beam $\gamma_{recoil} \sim 10~kHz$ \cite{jain_direct_2016}. The onset of thermos-optic instability prevented us from further increasing the laser pump power. This is presumed to be due to the contamination of the cavity used in the experiment. However, we note that stable operation of similar silica toroidal microcavities with the pump power above $2~W$ has previously been reported \cite{delhaye_octave_2011}. Therefore, we anticipate that moderate improvement of the device quality would allow for increasing the pump power up to a few milliwatts, thus further increasing the coupling. For instance, pump power of $2~mW$ would result in the coupling $g/2\pi \sim 450~kHz$. This enhancement would yield $C_q \sim 15$ at a pressure below $10^{-7}~mbar$, far exceeding the value previously reported in the field \cite{delic_levitated_2020, piotrowski_simultaneous_2023, delic_cooling_2020}.

An important aspect of our system is that the cavity exhibits exceptionally low optical loss. It allowed us to achieve the cavity loss to the particle frequency ratio of $\kappa/\Omega_z=11$, which is three orders of magnitude smaller than the previous work with photonic crystal microcavities \cite{magrini_near-field_2018}. This value is already enough to perform available protocols in the resolved-sideband regime, albeit with compromised fidelity. For instance, the achievable final thermal excitation of the particle’s motion by the sideband cooling\cite{wilson-rae_theory_2007, marquardt_quantum_2007} is $n_f \sim \kappa⁄4 \Omega = 2.8$. We note that the excess laser phase noise, which can substantially deteriorate the sideband cooling performance \cite{meyer_resolved-sideband_2019}, is not considered here as it can be suppressed in principle by using filter cavities \cite{hald_filter_cavity_2005}. The demonstrated value of $\kappa/2\pi$ ($4.4~MHz$), thus $\kappa/\Omega_z$, was limited by the fabrication imperfection and the contamination thereafter. However, it can be further improved to $\kappa⁄2\pi = 1~MHz$ as previously demonstrated \cite{kippenberg_demonstration_2004}. This improvement will not only allow us to achieve the sideband cooling near the quantum ground state ($n_f \sim 0.4$) but also to utilize various quantum protocols like coherent optomechanical state transfer \cite{wang_using_2012, palomaki_coherent_2013} and optomechanical entanglement \cite{hofer_quantum_2011, palomaki_entangling_2013} at the quantum regime.

So far, we have mainly discussed using our system in the context of standard cavity optomechanics, where the cavity is externally driven by the pump laser. We also expect our system to show enhanced performance in experiments based on coherent scattering \cite{delic_cs_2019, windey_cs_2019}. Recently, experiments based on coherent scattering have used macroscopic cavities to demonstrate control of levitated particles near the quantum ground state \cite{delic_cooling_2020, piotrowski_simultaneous_2023} and to achieve the strong coupling regime ($g>\kappa/4$) with projected cooperativities up to around 40 \cite{de_los_rios_sommer_strong_2021, dare_linear_2023}. Similar to standard cavity optomechanics, the coupling strength in coherent scattering ($g_{cs}$) also strongly depends on the cavity's mode volume (V), i.e., $g_{cs} \propto 1/\sqrt{V}$ \cite{delic_cooling_2020, piotrowski_simultaneous_2023}. 
Given the enhancements achieved in our experiment, we believe that our system could also offer substantial improvements in coupling strength and cooperativity with coherent scattering and reach the strong coupling regime.


\section{Funding Sources}
This research was supported by the Center for Integrated Quantum Science and Technology (IQST, Johannes-Kepler Grant) through the Carl Zeiss Foundation. Hansuek Lee acknowledges the support by National Research Council of Science and Technology (NST) (CAP21031-200) and KAIST Cross-Generation Collaborative Lab project. Haneul Lee acknowledges the support by the education and training program of the Quantum Information Research Support Center funded through the National research foundation of Korea (NRF) (2021M3H3A1036573).

\begin{acknowledgement}

We thank Prof. Harald Giessen for his support in the early stages of this research.

\end{acknowledgement}

\bibliography{ReferencesMain_v2}



\section*{Supplementary material}

# Supporting Information:

## Fabrication of microcavity

The toroidal microcavities are fabricated by the process shown in Fig. S 1a.[1] First, a 2 $\mu m$ thick silicon oxide layer is thermally grown from a [100] oriented silicon wafer (resistivity $1 \sim 10$ $\Omega$, p-doped). Subsequently, an array of 200 $\mu m$ diameter photoresist disks is created on the wafer by photolithography. Using them as etch masks, silica wedge disks are etched with buffered oxide etchant. The chip is then diced so that the silica disks are within 100 $\mu m$ from the chip edge. This later allows the optical tweezers to access the toroid's circumference from the side. After dicing, the remaining photoresist and contaminants are removed with an organic remover. The silicon substrate is isotopically etched using a $XeF_2$ dry etch system, creating a silicon post to thermally isolate the silica disks from the silicon substrate during $CO_2$ laser reflow. Finally, the silica disks are thermally reflowed into a toroid with an ultra-smooth surface by applying a focused $CO_2$ laser pulse (10.6 $\mu m$ wavelength, $P < 10$ $W$), which selectively heats the silica layer.[2,3] The scanning electron microscopy (SEM) images of a single toroidal microcavity are shown in Fig. S 1b and c.

In the $XeF_2$ etching process, it is important to ensure that undercut silicon posts are formed radially symmetric. Failure to do so could result in asymmetric heat flow through the posts during the reflow process, adversely affecting the shape of the toroid and the resulting Q factor. In this study, the chips are diced so that the silica disks are within 100 $\mu m$ from the diced edge (Fig. S 2a). This results in the formation of the silicon post shifted to the opposite side of the chip edge from the disk center. This is because the part closer to the chip

edge is more exposed to the $XeF_2$ gas during the etching process (Fig. S 2c). We resolve this issue by placing a dummy Si chip right next to the chip edge (Fig. S 2d), resulting in a formation of the silicon post at the disk center (Fig. S 2b).

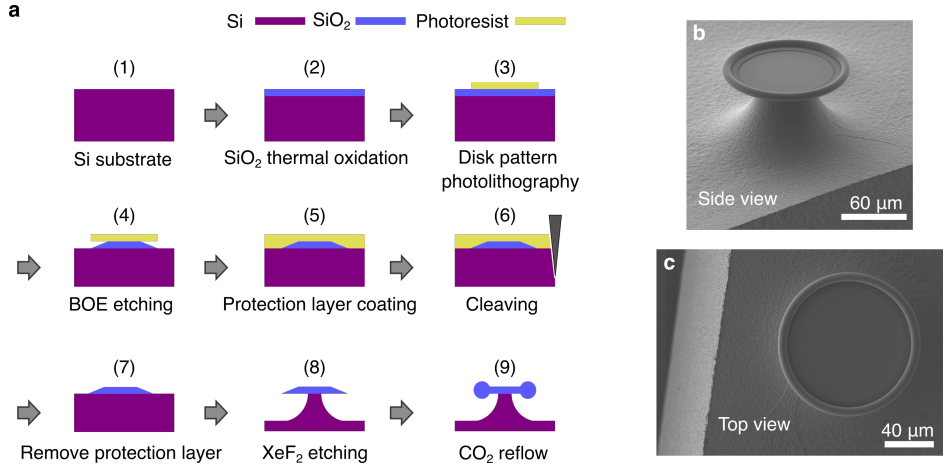

**Fig. S1: a** Schematic illustration of fabrication process of a toroidal microcavity. Here the view is from the side. **b** and **c**, Scanning electron micrographs of a single microcavity. The view is from the side in b and top in c. The microcavity is fabricated within $100 \mu m$ from the edge of the substrate which allows it to be brought towards the focal spot of the horizontally-aligned optical tweezer.

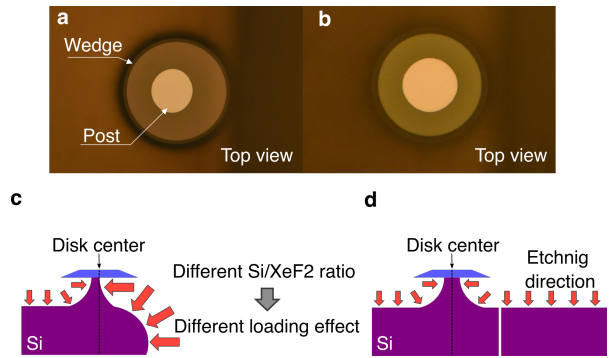

**Fig. S2: a** and **b**, Optical microscope images of the microcavities from a top view. The $XeF_2$ etching process results in non-uniform etching of the Si substrate and an off-center post as shown in a. To prevent non-uniform etching, a dummy Si piece is employed, resulting in uniform etching shown in b. **c** and **d**, depict a schematic of the etching process of the microcavities without and with a dummy Si piece, respectively.

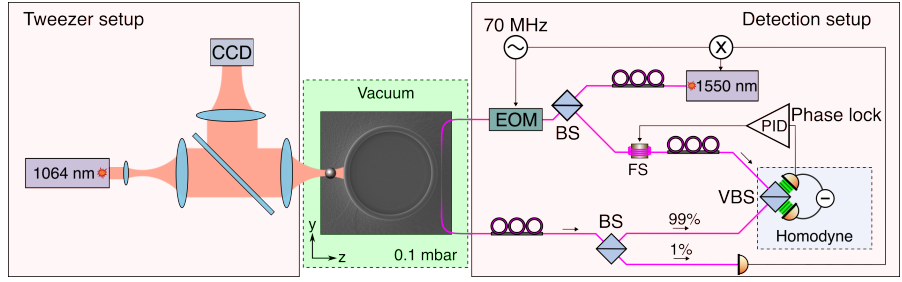

**Fig. S3:** Schematic of the experimental setup. The left part shows a standard tweezer setup for the levitation of a single nanoparticle. This tweezer is created by tightly focusing a 1064nm laser light through a 0.95 NA microscope objective. A CCD camera is employed to obtain the image of the trapped nanoparticle and the microtoroid. Shown on the right side is a fiber-based setup used for pumping the microcavity and performing homodyne detection of the cavity output field. Further details of this setup are provided in section 'Experimental setup'.

## Experimental setup

Figure S 3 shows an overview of the experimental setup. A tunable fiber laser (Toptica CTL1550; 1500 $nm \sim$ 1600 $nm$) is used to pump a single microcavity. Initially, the laser light is split into a high- and a low-power paths. The high-power beam serves as a local oscillator (LO), while the low-power beam is evanescently coupled to the microcavity using a tapered fiber (see Ref. 4 and 5 for further details). The transmission of the cavity is collected with the same fiber. This signal is split at a 99:1 ratio for homodyne detection and for locking the pump laser to the cavity resonance.

In the homodyne path, the cavity response is interfered with the LO using a variable ratio coupler set to a 50:50 ratio. The combined signals are then directed to a balanced photodetector. The LO is phase-locked to the cavity output using a fiber piezo stretcher. At the same time, the 1 % part of the cavity transmission is detected using a variable gain photodetector and used for locking the pump laser. We utilize the Pound-Drever-Hall technique for laser locking. To implement this method, we modulate the phase of the laser input to the cavity using fiber-coupled electro optic modulator (EOM) at the frequency of 70 $MHz$. This modulation induces sidebands in the cavity transmission when the laser frequency is scanned around the resonance (see also Fig. S 5b). The resultant signal,

obtained by measuring the transmission, is mixed with the modulation signal. This mixed signal generates an error signal for locking the laser.

# Increase in particle's oscillation frequency

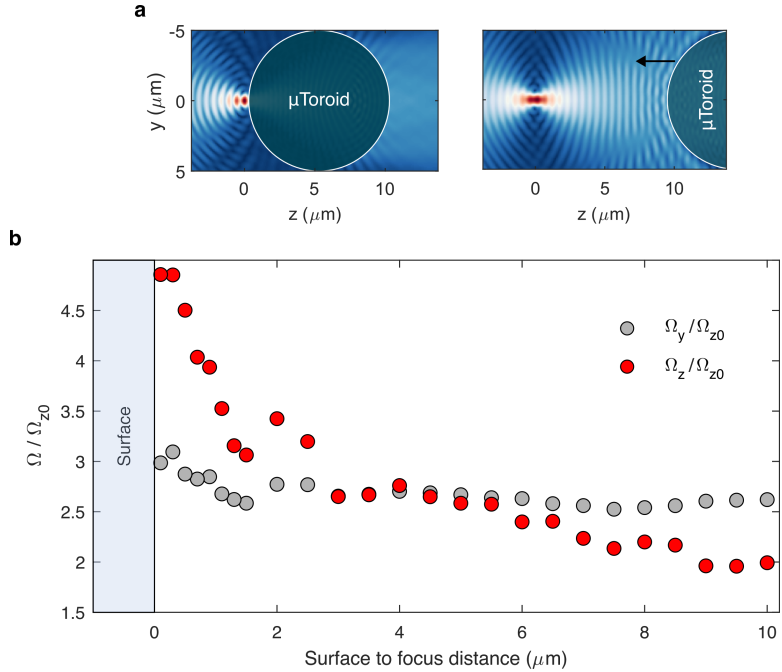

**Fig. S4: a**, Simulation of standing waves and potential wells created by the reflection of the tweezer laser field by the microcavity surface. On the right (left) panel, the modified amplitude of the tweezer electric field is shown when the microcavity is placed $10\mu m$ ($300nm$) away from the focal plane ($z = 0$). Moving the microcavity closer to the focal spot results in a stronger localization of the reflected beam and increases the stiffness of the trap, leading to an increase in the oscillation frequency of the particle.[6] **b**, The normalized frequency of the particle's motions along the tweezer $z$-axis ($\Omega_z$) and for a transverse direction ($\Omega_y$) as a function of the distance from the surface of the cavity to the focal point of the tweezer beam. The frequencies are normalized to $\Omega_{z0}$, which is the frequency of the $z$-motion imposed solely by the tweezer beam in the absence of the microcavity. Notably, at short distances, A significant increase of $\Omega_z$ is observed, while $\Omega_y$ remains relatively unchanged. The trend of $\Omega_x$ (motion along the $x$-axis) follows a pattern similar to that of $\Omega_y$.

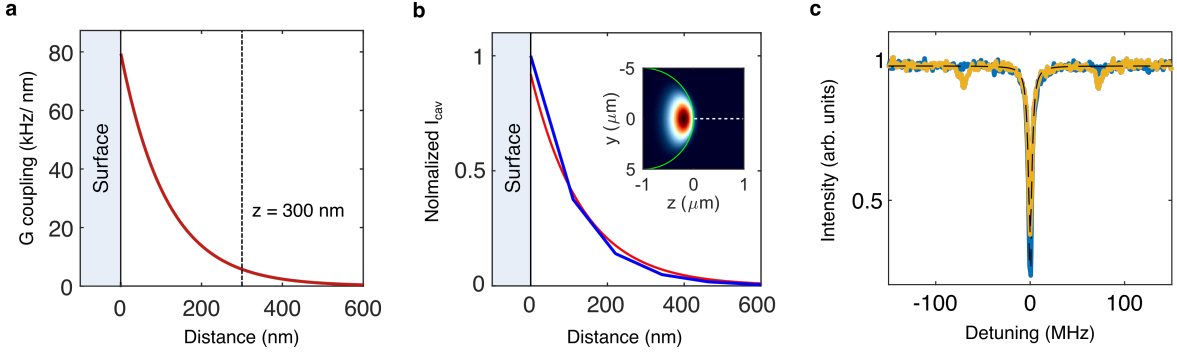

**Fig. S5: a,** Simulation of the cavity frequency shift per particle displacement caused by the center of mass (COM) motion of a single silica nanoparticle of 150 $nm$ diameter. The frequency shift per displacement exponentially increases by approaching the cavity surface in accordance with the evanescent field shown in b. The dashed line shows the position of the closest potential well, i.e., 300 $nm$, created by the formation of the standing wave, where $G = 5.8\ kHz/nm$. **b** The intensity profile of the evanescent field along $z$-axis for the fundamental mode of the microcavity extracted from the simulation. The inset shows the 2D map of this mode in the y-z plane. The dashed line indicates the line on the equator plane, where the intensity profile is extracted. The red line is an exponential fit to the data with a decay length of 133.5 $nm$. As expected from Eq. 1, $G$ is proportional to the intensity of the evanescent field. **c,** An example of the cavity transmission when the particle is coupled to the fundamental mode of the cavity (at a resonance of $\lambda \approx 1550.06\ nm$) as a function of the laser detuning. To generate the error signal for the PDH locking technique, the cavity is pumped with a phase-modulated laser beam. This results in sidebands at $\pm 70\ MHz$ (yellow) in the transmission of the cavity when the pump laser is scanned around the resonance.

## Optomechanical coupling

### Simulation:

To estimate the optomechanical coupling between the particle and the cavity field, we performed a finite element method (FEM) simulation. The presence of a small dielectric particle at the position of $z$ perturbs the environment of the microcavity and results in a frequency shift of,[7]

$$\Delta\omega(z) \approx -\frac{\omega_0}{2}\frac{\int_{V_{np}}(\epsilon_{np}-1)^2|E_0(\vec{r})|^2\,d\vec{r}}{\int_{V_{cav}}|E_0(\vec{r})|^2\,d\vec{r}}. \quad (1)$$

Here, $E_0(\vec{r})$ is the electric field amplitude of the cavity mode with the resonance frequency

of $\omega_0$ in the absence of the nanoparticle and $\epsilon_{np}$ is the relative dielectric constant of the nanoparticle. The integration is performed on the volume of nanoparticle ($V_{np}$) located in $z$, and the mode volume of the cavity ($V_{cav}$). The linearized coupling between the particle and the cavity can be extracted as $G = \frac{\partial \omega}{\partial z}\big|_{z=q}$ where $q$ is the position of the nanoparticle. Figure S 5a shows the G coupling as a function of the distance from the surface of the cavity. This coupling is expected to be 5.8 $kHz/nm$ at the distance of 300 $nm$ (dashed line) for a microtoroid with the major and minor diameters of 80 $\mu m$ and 7 $\mu m$, respectively. In the field of quantum optomechanics, it is a convention to express the optomechanical coupling $g_0$ in units of zero-point fluctuation of the mechanical part $x_{zpf} = (\hbar/2m\Omega)^{1/2}$, i.e., $g_0 = G \cdot x_{zpf}$.[8] We follow this convention here and throughout the main text. The optomechanical coupling can be obtained by measuring $G$ and calculating $x_{zpf}$ with known values of particle mass and frequency.

## Measurement:

Here we describe in detail how we measure $G$ experimentally.

When a cavity is pumped by an input field $\hat{a}_{\text{in}}$, the rate at which the photon arrives at the cavity is represented by $\langle \hat{a}_{in}^\dagger \hat{a}_{in} \rangle$, and the input power launched into the cavity is expressed as $P_{in} = \hbar \omega_{cav} \langle \hat{a}_{in}^\dagger \hat{a}_{in} \rangle$. Inside the cavity, the photons decay at a rate of $\kappa = \kappa_{ex} + \kappa_0$, which includes both the decay due to the input-output coupling $\kappa_{ex}$ and the intrinsic cavity losses $\kappa_0$. This system can be described by means of the input-output theory,[8] where the output field $\hat{a}_{\text{out}}(t)$ is related to the input field and the field of the cavity mode $\hat{a}(t)$ as follows:

$$\hat{a}_{\text{out}}(t) = \hat{a}_{\text{in}}(t) - \sqrt{\kappa_{ex}}\hat{a}(t), \tag{2}$$

where the cavity field is given by,

$$\hat{a} = \frac{\sqrt{\kappa_{ex}} \cdot \hat{a}_{in} + \sqrt{k_0} \cdot \hat{a}_0}{-i\Delta + k/2}. \tag{3}$$

Here, $\hat{a}_0$ is the vacuum fluctuation field amplitude, $\Delta = \omega_l - \omega_{cav}$ is the detuning of the pump field frequency $\omega_l$ to the cavity resonance $\omega_{cav}$. Here we assume that the cavity is pumped at resonance ($\omega_l = \omega_{cav}$), and we note that the presence of a mechanical oscillator coupled to the cavity field will induce frequency fluctuations in the cavity resonance, such that $\omega_{cav} \to \omega_{cav} + \delta\omega_m$. Substituting Eq. 3 into Eq. 2, we write

$$\hat{a}_{\text{out}}(t) = \hat{a}_{in} - (\kappa_{ex}\hat{a}_{in} + \sqrt{\kappa_{ex}\kappa_0}\hat{a}_0)\left(\frac{k/2}{\delta\omega_m^2 + (k/2)^2} + i\frac{\delta\omega_m}{\delta\omega_m^2 + (k/2)^2}\right). \quad (4)$$

For simplicity, we assume $\hat{a}_{in}(t)$ is a strong coherent drive field plus small fluctuations, expressed as $\alpha_{in} + \delta\hat{a}_{in}(t)$. Here $\alpha_{in}$ is the mean amplitude of the field with a real value and $\delta\hat{a}$ represents small fluctuations around this mean. Using Eq. 4, we derive the output field of the cavity for such an input. The phase quadrature of the output field is $\delta Y = \frac{i}{\sqrt{2}}\left(\delta\hat{a}_{\text{out}} - \delta\hat{a}^\dagger_{\text{out}}\right)$ and its associated phase spectral density, $S_{YY}(\Omega)$, is given by,

$$S_{YY}(\Omega) = \int_{-\infty}^{\infty} e^{-i\omega t}\langle \delta Y(t)\delta Y(0)\rangle \, dt. \quad (5)$$

Upon substituting Eq. 4 into Eq. 5, we obtain the relation expressing the phase spectral density in terms of the spectral density of the cavity's frequency in the classical limit:

$$S_{YY}(\Omega) = 2\frac{4\eta_c^2 \alpha_{in}^2}{\kappa^2} S_{\omega\omega}(\Omega). \quad (6)$$

This indicates that by measuring the phase quadrature of the cavity response, its frequency fluctuations can be resolved. Here, $\eta_c = \frac{2\kappa_{ex}}{\kappa}$ is introduced to account for deviation from the critical coupling, where $\kappa_{ex} = \kappa_0 = \kappa/2$. For the measurement results presented in Fig. 2e and 3 in the main text, $\eta_c = 0.8$.

On the other hand, we recall that frequency fluctuations are induced by the particle displacement, i.e,

$$S_{\omega\omega}(\Omega) = G^2 S_{xx}(\Omega) \tag{7}$$

The $S_{xx}(\Omega)$ is related to variance of the particle $\langle x^2 \rangle$, where,

$$\langle x^2 \rangle = \int_{-\infty}^{\infty} S_{xx}(\Omega) \frac{d\Omega}{2\pi}. \tag{8}$$

When the particle is in thermal equilibrium with its surroundings, $\langle x^2 \rangle = \frac{k_B T}{m\Omega_m^2}$. The integration of Eq. 7 leads to the following equation:

$$G = \sqrt{\frac{m\Omega_m^2}{k_B T} \int_{-\infty}^{\infty} S_{\omega\omega}(\Omega) \frac{d\Omega}{2\pi}} = \sqrt{\frac{m\Omega_m^2}{k_B T} \cdot \frac{\kappa^2}{8\eta_c^2 \alpha_{in}^2} \int_{-\infty}^{\infty} S_{YY}(\Omega) \frac{d\Omega}{2\pi}}. \tag{9}$$

Thus, we learn that measurement of $S_{YY}(\Omega)$ can obtain $G$.

$S_{YY}(\Omega)$ can be measured by performing the **homodyne measurement**. In this method, the cavity response, driven with an input power of $P_{in}$, is interfered with a strong local oscillator ($P_{lo}$) and consequently sent to a (50:50) balanced photodetector. By selecting the optimal phase difference between the signal and the LO, we can record the phase quadrature of the cavity response. The resulting photocurrents from both detectors are subtracted and subsequently converted to voltage and amplified by trans-impedance gain of $G_t$, yielding to:

$$S_{VV}(\Omega) = 2G_t^2 \eta_{loss} R^2 P_{LO} \hbar \omega_{cav} S_{YY}(\Omega). \tag{10}$$

This voltage is then recorded by an oscilloscope. Here, $\eta_{loss} = 0.17$ represents the total loss in the signal path, including cavity output $\eta_{cav} = 0.81$ and all fiber-to-fiber losses in the setup. $R = \frac{\eta_q e}{h\nu}$ represents the responsivity, $\eta_q = 0.85$ is the quantum efficiency of the detector, and $e$ denotes the electron charge.

# Steering measurement

In the steering measurements, the microtoroid is first moved towards the objective (along $z$-axis) in such the particle is placed in the closet potential well and its motion is coupled to the cavity. While the particle is kept in the closest well, we steer the tweezer beam along $y$-axis. The CCD camera installed in the tweezer setup provides information about the position of the particle while steering the beam. Examples of its recorded images are shown in Fig. S 6c. These images are processed to extract the location of the trap beam. The coordinates on the images are calibrated using a reference microsctucture. For each selected location, we collected time traces of the cavity's homodyne signal over a duration of $\approx 600\ ms$. These time traces were segmented into ten snippets, each approximately $60\ ms$ long. Then the power spectral density (PSD) for each snippet is calculated and averaged to obtain one PSD for each measurement. The averaged PSD is used to calculate $S_{YY}$ via Eq. 10 and subsequently to extract the values of the optomechanical coupling, G, using Eq. 9 and thus $g_0$ and $g$. This process was repeated 30 times over a total duration of about 30 seconds (including delays between measurements and data transfer time). Each data point in Fig. 3c represents the mean value of g, averaged over these 30 measurements, and the error bars indicate the deviation from its mean value. The measured PSDs corresponding to the data presented in Fig. 3c and d are shown in Fig. S 6b.

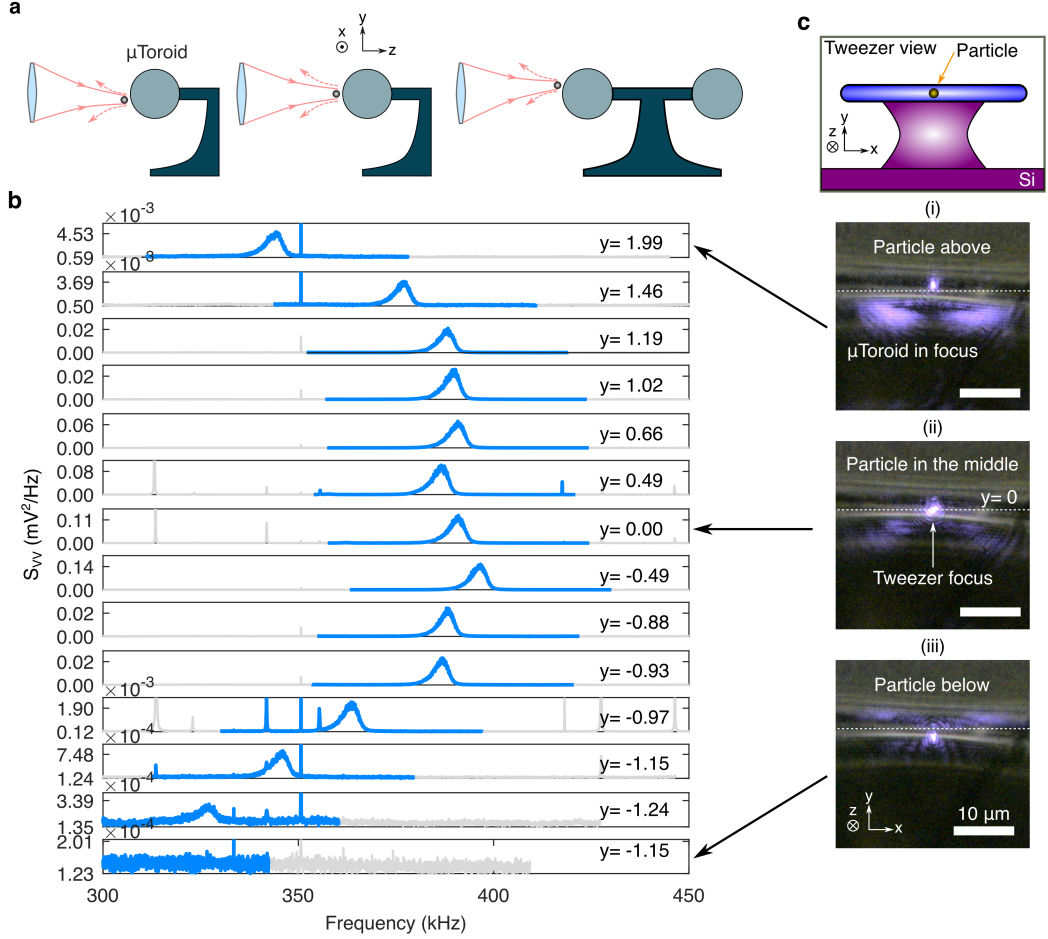

**Fig. S6:** Tuning of optomechanical coupling and particle's oscillation frequency. **a**, Illustration of beam steering with a cross-sectional profile of the microtoroid. The tweezer beam is steered along the $y$-axis with the particle positioned in the closest potential well. Simultaneously, the cavity's response is recorded through homodyne measurement. **b**, The PSDs of the measured homodyne signal during the steering measurement. To enhance data visibility, the $y$-axis scale is adjusted for each subplot. The legends show the beam position for each measurement in the units of $\mu m$. While the tweezer beam is steered along $y$-axis, its position is tracked using a CCD camera installed in the tweezer setup (shown in Fig. S3). Camera images are processed to extract the beam's position. **c** Schematic of the tweezer view and examples of recorded CCD images. The dashed line shows the original position (middle) of the tweezer beam, $y = 0$ (in panel (ii)). The particle is steered above and below the middle line in panels (i) and (iii), respectively.

# Optomechanical coupling to microcavity's higher modes

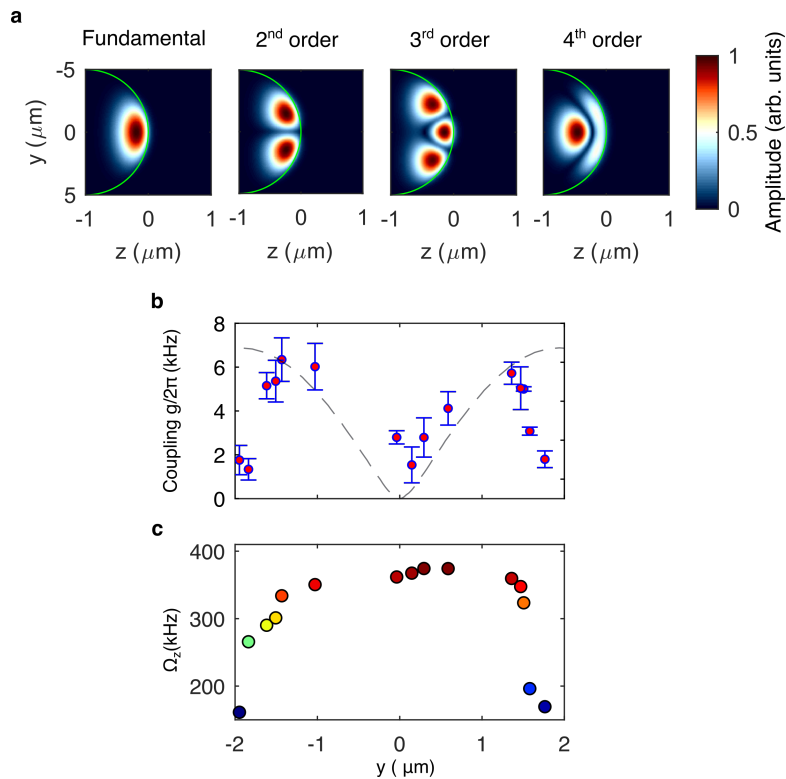

**Fig. S7: a**, Illustration of the profiles of the fundamental and higher-order modes of the microtoroidal cavity, each displaying distinct spatial and spectral characteristics. **b**, Measured optomechanical coupling of the COM motion of a particle to the second-order microcavity mode. The mode exhibits resonance at a wavelength of $\lambda = 1553.22\ nm$ with $\kappa/2\pi = 6.6\ MHz$. In contrast to the fundamental mode, the coupling strength diminishes at the equator plane ($y = 0$), following the profile of the second-order cavity mode (depicted in a). The dashed line represents the profile of the second mode, serving as a visual guide. By moving away from the equator plane, the trapping beam becomes deformed, and the preservation of the particle-to-surface distance is no longer maintained, resulting in a rapid reduction in coupling strength. **c**, Frequency of the particl's $z$-motion for the same measurement.



\end{document}